\documentstyle[preprint,aps,floats,epsf,rotating,epsfig]{revtex}
\begin{document}
\def\beq{\begin{equation}}
\def\eeq{\end{equation}}
\def\ov{\overline}
\thispagestyle{empty}
\draft
%\input{first_page}
%\starttext
\pagestyle{columns}
%\begin{titlepage}
\input{first_page}
%\newpage
%\def\slash#1{#1 \hskip -0.5em / }
%***************************************************************************
\begin{flushright}CEBAF-TH-94-24\end{flushright}\vspace{2.0 cm}

\begin{center}
BOUNDS ON $\Delta B=1$ COUPLINGS IN  THE SUPERSYMMETRIC STANDARD MODEL

\vskip 0.5in

J. L. Goity

\vskip 0.2in

{\it Department of Physics, Hampton University, Hampton, VA 23668, USA

and

Continuous Electron Beam Accelerator Facility

12000 Jefferson Avenue, Newport News, VA 23606, USA.}

\vskip 0.2in

 Marc. Sher
\vskip 0.2in

{\it Department of Physics,  College of William and Mary,
%Williamsburg, VA 23187, USA.}
\end{center}

\thispagestyle{empty}
\title{BOUNDS ON $\Delta B=1$ COUPLINGS IN  THE  SUPERSYMMETRIC STANDARD MODEL}
%\vspace*{1.cm}\\
\author{J. L. Goity}
\address{
Department of Physics, Hampton University, Hampton, VA 23668, USA \\
and \\
Continuous Electron Beam Accelerator Facility \\
12000 Jefferson Avenue, Newport News, VA 23606, USA.}

\author{
Marc Sher}
%}
\address{
Department of Physics,  College of William and Mary,
Williamsburg, VA 23187, USA }
%\date{\today}
\maketitle
\begin{abstract}
 The most general supersymmetric model contains baryon number violating terms
of the form $\lambda_{ijk}\;\ov{D}_i\, \ov{D}_j\, \ov{U}_k$ in the
superpotential.
We reconsider the bounds on these couplings, assuming that lepton number
conservation ensures proton stability.  These operators can mediate
$n -
\ov{n}$ oscillations and double nucleon decay.  We show that neutron
oscillations do not, as previously claimed, constrain the $\lambda_{dsu}$
coupling; they do provide a bound on the $\lambda_{dbu}$ coupling, which we
calculate.  We find that the best bound on $\lambda_{dsu}$ arises from double
nucleon decay into two kaons; the calculation is discussed in detail.
There are no published limits on this process; experimenters are urged
to examine this nuclear decay mode.  Finally, the other couplings can be
bounded
by the requirement of perturbative unification.

\end{abstract}
%\pacs{  {\tt$\backslash$\string pacs\{13.25.Hw, 1.39.Fe, 14.40.Lb, 14.40.Nd
%%\}} }
%\end{titlepage}
\thispagestyle{empty}
\newpage
\setcounter{page}{1}

In the standard electroweak model, conservation of baryon number and
lepton number arises automatically from gauge invariance.  This is not
the case in supersymmetric models, however.  In the most general
low-energy supersymmetric model, one has terms which violate lepton
number and terms which violate baryon number\cite{bviol}.  Since the
presence of both of these may lead to unacceptably rapid proton decay
(unless the couplings are extraordinarily small), one or both must
generally be suppressed by a discrete symmetry\cite{symm}. In the most
popular model, $R$-parity, given by
$(-)^{3B+L+F}$, where $B$, $L$ and $F$ are the baryon number, lepton
number and  fermion number, is imposed, leading to baryon and lepton
number conservation. However, there is no {\it a priori} reason that
$R$-parity must be imposed (other than a desire to conserve baryon
number, lepton number and to simplify phenomenology); it is quite
possible that only one of the quantum numbers is conserved.  There has
been extensive discussion of the possibility that lepton number is
violated\cite{lepton}, but relatively little investigation of the
possibility that lepton number is conserved and baryon number is
violated.

In this case, baryon number conservation will be violated in the
low energy theory. A term will appear in the superpotential given by
\beq\lambda_{ijk}\ov{D}_i\ov{D}_j\ov{U}_k\big|_F\eeq where the indices
give the generation number and the chiral superfields are all right-handed
isosinglet antiquarks. Since the term is symmetric under exchange of the
first two indices, and is antisymmetric in color, it must be antisymmetric
in the first two flavor indices, leaving nine couplings, which
will be designated (in an obvious notation)
\def\la{\lambda}
as $\la_{dsu},\ \la_{dbu},\ \la_{sbu},\
\la_{dsc},\ \la_{dbc},\ \la_{sbc}$,  $\ \la_{dst}$, $\ \la_{dbt}, {\rm
and}\  \la_{sbt}$.

There are many models in which low energy baryon number is violated
and yet lepton number is conserved; the most familiar are some
left-right symmetric models (see Ref.\cite{rabi} for a comprehensive
discussion).  It is widely believed that the strongest bound on
B-violating operators comes from neutron oscillations, which
violate $B$ by two units.  The first discussion\cite{zwirner} on the
effects  of some of these operators in supersymmetric models used neutron
oscillations to bound
the $\la_{dsu}$ and $\la_{dbu}$ couplings; this result remains
widely cited today\cite{zwcite,roy}.

The purpose of this Letter is to determine the most stringent bounds
that can be placed on these nine couplings.  First, we will point out
that neutron oscillations do {\it not} provide any significant bound
at all on the $\la_{dsu}$ coupling, due to a suppression factor which
was neglected in the original calculation.  This suppression is  less
  severe for the $\la_{dbu}$ coupling, and we will obtain a
bound in that case.  We will then note  that the
strongest bound on the $\la_{dsu}$ coupling will come from limits on
double nucleon decay (in a nucleus) into two kaons of identical
strangeness, and will estimate the bound.  A recent work of
Brahmachari and Roy\cite{roy} noted that the $\la_{dbt}$ and $\la_{sbt}$
couplings can be bounded by requiring perturbative unification; we
will extend their work to cover all of the additional couplings.
Finally, we will comment on various bounds which can be obtained by
considering products of some of the couplings.\footnote{Severe bounds on some
of the couplings have been suggested from cosmological arguments, however,
Dreiner and Ross\cite{zwcite} have shown that these bounds can be evaded.}

The first attempt to place a bound on some of the above operators
was in Ref. [5].  There, the contribution of the $\la_{dsu}$
and $\la_{dbu}$ terms to neutron oscillations was calculated by
considering the process
$(udd\rightarrow \ov{\tilde{d}}_i\, d\rightarrow \tilde{g} \rightarrow
\tilde{d}_i\, \ov{d}\rightarrow \ov{u}\, \ov{d}\, \ov{d}$, where $\tilde{d}$ is
a
squark and $\tilde{g}$ is a gluino.  The effects of intergenerational mixing
were included by putting in an arbitrary mixing angle, assigned a value of
$0.1$. However, there is a much more severe suppression factor which
results in this process giving no significant contribution to neutron
oscillations.

Consider the term $\la_{dsu}\ov{D}\ \ov{S}\ \ov{U}$.  It
violates $B$ by one unit, and it violates strangeness (S) by one
unit.  However, it conserves $B-S$.  Since neutron oscillations
violate $B$ but {\it not} $S$, strangeness violation must appear
somewhere else in the diagram.  This means that there must be
flavor-changing electroweak interactions (involving either a  $W$ or
a charged
Higgs  and charginos) in the diagram\footnote{Should additional
sources of strangeness violation exist, such as tree-level
flavor-changing neutral currents due to an extended Higgs sector, then
electroweak interactions would be unnecessary.}. This flavor change is produced
by the box diagram of Figure 1.   One can also replace the
$W$ with a charged Higgs boson, and the $\tilde{W}$ should be the
lightest chargino, but we will consider the case in which the
charged Higgs boson is much heavier than the $W$, and there is
little mixing in the chargino sector; for an order of magnitude
estimate, this will be sufficient. Since only isodoublets
participate in the  weak interactions, and the baryon number violating term
only has isosinglets,
there must be  mass insertions (at least two, in the simplest case) which
give a transition from $\tilde{d_R}_i$ to  $\tilde{d_L}_i$. This mass insertion
will be proportional to the mass of the
associated quark. Thus, there will be electroweak interactions in the
diagram, and an additional suppression factor of the order of
$m_s^2/m^2_W$, where $m_s$ is the strange quark mass. This makes the
contribution of the $\la_{dsu}$ highly suppressed.

The contribution involving the term $\la_{dbu}\ov{D}\,\ov{B}\,\ov{U}$ will be
suppressed by a
factor of $m_b^2/m^2_W$, which is not negligible, and this
could give a significant contribution to neutron oscillations.  The
relevant diagram is given in Figure 2, where the  flavor changing  box
subdiagram
is represented by a blob;  the two  possible contractions of the legs of the
box   give an
equal contribution. The resulting dimension nine effective operator can be
written as (greek indices correspond to color)
\beq T\;\epsilon_{\alpha\beta\gamma}\,\epsilon_{\alpha'\beta'\gamma'}\;
\ov{u}^c_{R\alpha}
d_{R\beta}\;\ov{d}^c_{L\gamma}d_{L\gamma'}\;\ov{u}^c_{R\alpha'}d_{R\beta'}
\eeq
The  diagram is calculated at zero external momenta and yields
\beq T=-{3g^4\la_{dbu}^2 M_{\tilde{b}_{LR}}^2 m_{\tilde{w}}\over
8\pi^2M^4_{\tilde{b}_L}
M^4_{\tilde{b}_R}}\xi_{jj'}J(M^2_{\tilde{w}},M^2_W,M^2_{u_j},M^2_{\tilde{u}_{j'}})\eeq
where the mass term $M_{\tilde{b}_{LR}}$ which mixes   $\tilde{b}_{L}$ and
$\tilde{b}_{R}$
is given by $M_{\tilde{b}_{LR}}=A\,m_{b}$,   $A$ is the soft supersymmetry
breaking parameter  (there is also an
F-term contribution which we absorb into $A$); $j$ and $j'$ are generation
indices, $\xi_{jj'}$ is a combination of KM angles (we assume the
left-handed squark KM matrix is the same as that of the quarks) :
\beq\xi_{jj'}=V_{bu_j}V^{\dagger}_{u_jd}V_{bu_{j'}}V^{\dagger}_{u_{j'} d}
\eeq
and \beq J(m^2_1,m^2_2,m^2_3,m^2_4)=\sum_{i=1}^4{m^4_i\ln(m^2_i)\over
\prod_{k\ne i}(m^2_i-m^2_k)}\eeq
The neutron oscillation time is then given by $\tau=1/\Gamma$, where
$\Gamma=T\psi(0)^2$.  $\psi(0)^2$  gives the matrix element of the
  operator in (2);  we use the
estimate given by Pasupathy\cite{pasu}, $\psi(0)^2=  3\times 10^{-4}$
GeV${}^6$,
but it should be noted that other evaluations\cite{pasuothers} differ by
more than an order of magnitude (the bound on $\la_{dbu}$ will vary as the
square root of
$\psi(0)^2$).  From the experimental limit on the neutron oscillation
time\cite{pdg}, $\tau\ >\ 1.2\times 10^8$ sec., we can obtain the bound
on $\la_{dbu}$.  The results depend on the Kobayashi-Maskawa angles,
which are taken to be the central values of the allowed ranges, and the
squark masses.  It is assumed, as is the case in most models, that the
charm and up squark masses are degenerate.  The bound is plotted in Figure
3 as a function of the top squark mass for various   charm squark masses. We
keep $A=m_{\tilde{w}}=200$ GeV throughout. Note the peaks which
correspond to GIM cancellations in the box diagram.  Unless the
parameters are tuned to this cancellation, the upper bound on
$\la_{dbu}$ is between $0.002$ and $0.1$ if the squark masses are between 200
and 600 GeV, with more stringent limits resulting for
lighter masses.  As stated above, the bound
on $\la_{dsu}$ will be weaker by roughly a factor of $m_b/m_s$, and will
not be competitive with the bound in the following paragraph. The bound on the
square root of the product $|\la_{dsu}\la_{dbu}|$ is suppressed instead
by a factor $\sqrt{m_b/m_s}$.
As  Figure 3 suggests,  unless the supersymmetric particles are lighter than
1 TeV, no useful bounds result from $n-\ov{n}$ oscillations.
In fact, better bounds can then  be derived from perturbative unification as we
 discuss
later.

The fact that the best bound on $\la_{dsu}$ comes from double
nucleon decay into two kaons of identical strangeness was noted some
time ago\cite{barb,hall}.  It is easy to see that the mass insertions and
electroweak interactions are unnecessary, since the process does violate
both B and S by two units.  In Ref. 12, a rough order of magnitude
estimate was given for the bound (with no details of the estimate).  In
Ref. 11, it was assumed that, in a nucleus, a neutron could ``oscillate"
into a
$\ov{\Xi}$ through two applications of the operator, and that the
 $\ov{\Xi}$ annihilates with
another neutron in the nucleus to produce two kaons.  Here, we will
consider the process $NN\rightarrow KK$ directly. The relevant diagram is
shown in Figure 4, which reduces to  dimension
nine operators of the form:
\begin{equation}
\frac{16}{3}
\;g_s^2\;\lambda _{dsu}^2
\frac{1}{M_{\tilde{g}} M_{\tilde{q}}^4}\;
\epsilon_{\alpha\beta\gamma} \,\epsilon_{\delta\rho\sigma}
\;\left( \overline{u^c_{R}}_\alpha d_{R\beta} \;
 \overline{u^c_{R}}_\delta d_{R\rho} \;
 \overline{s^c_{R}}_\gamma s_{R\sigma} +\cdots\right),
 \end{equation}
 where the ellipsis indicate all possible permutations between
the symbols $(u,d,s)$ and the symbols $(u^c,d^c,s^c)$.

The final state must contain  minus two units of strangeness,
 plus some pions. The strange  component of the
final state can  be any of the following:
 $K^{+,0} K^{+,0}$, $K^{+,0} \Lambda $, $ K^{+,0} \Sigma $.
Since for each possible final state   the  corresponding
amplitude
  contains a large number of possible arrangements of the quark
legs of the effective operators, a good estimate of the
individual rates is very difficult.  For the purpose of  giving a bound
on the order of magnitude of $| \lambda_{dsu} |$, a rough
estimate of those
rates should suffice, as they are
proportional to the fourth power of $| \lambda_{dsu} |$. We, therefore,
make the following simplyfying assumptions: a) The nine terms in (6) add up
roughly incoherently, and give
similar  contributions to the total rate; b) The  amplitudes for
individual rates are estimated using dimensional arguments, where the
relevant scale is a
  hadronic scale $\tilde{\Lambda}$, which we will let vary within a
generous range;
 c) It is sufficient for our purpose to consider only a few
final states, in particular,  the final state $KK$.
We have checked that the addition of
 other final states like  $KK\pi$ and $KK\pi\pi$ hardly affect
our  results.

 The total rate is given by:
 \begin{equation}
\overline{\Gamma}=\frac{1}{(2 \pi)^3 \rho_N}\;\int \;d^3 k_1
\;d^3
 k_2
\;\rho(k_1 ) \,\rho(k_2 )
 \;v_{rel} \,(1-\vec{v_1 }\cdot \vec{v_2 })\;\sigma_{tot}
(NN\rightarrow X),
 \end{equation}
where  $\rho_N$ is the average nucleon density, $\rho(k)$ is
the nucleon density   in momentum space,
 and the nucleon velocities are taken in the following as
small. Using our assumptions, the cross section is approximately
given by:
\begin{equation}
 \sigma_{tot} (NN\rightarrow X)\sim
\frac{128\,\pi\,\alpha_s^{2}\;|\lambda_{dsu}|^4\;
 C_{KK}}{v_{rel}\,M_N^2\;M_{\tilde{g}}^2\, M_{\tilde{q}}^8}.
\end{equation}
Here  the final state phase space was taken to be that for two
massless particles.
$C_{KK}$ has dimension ten and is  approximated by
$\tilde{\Lambda}^{10}$.   This scale is hard to estimate; direct annihilation
of the
two nucleons by the dimension nine operator is suppressed due to hard-core
repulsion, while
the contribution due to t-channel $\Xi$  exchange  may be the dominant piece.
With this we finally obtain the
following bound:
\begin{equation}
\overline{\Gamma}\sim  \rho_N \;
 \frac{128\,\pi\,\alpha_s^{2}\;\lambda_{dsu}^4\;
\tilde{\Lambda}^{10} }{M_N^2\;M_{\tilde{g}}^2\, M_{\tilde{q}}^8}.
\end{equation}
Using nuclear matter density $\rho_N=0.25$ $ {\rm fm^{-3}}$,
$\alpha_s\sim 0.12$ and a lower bound
for nuclear matter lifetime  of $\tau_N\sim 10^{30}$ years, we
obtain the following
 bound on $| \lambda_{dsu}|$
in terms of the ratio between the   hadronic  and the
supersymmetric scales
$R=
\frac{\tilde{\Lambda}}{(M_{\tilde{g}} \,
M_{\tilde{q}}^4)^{1/5}} $:
\begin{equation}
| \lambda_{dsu}| <  10^{-15}\; R^{-5/2}
  \end{equation}
 This ranges  from  as low as $10^{-7}$ for
$R\sim 10^{-3}$,
to   1 for $R\sim 10^{-6}$).
Our bound is comparable to  the bound obtained by Barbieri
and Masiero\cite{barb}  by
considering the transition
$N\rightarrow \ov{\Xi}$ in nuclei: they obtained a bound
approximately
 given by  $5\times 10^{-16} R^{-5/2}$.
Although their hadronic scale is not necessarily identical to ours,
we expect them to be similar in value.

What can be said about the other seven couplings?.   It was recently
noted by Brahmachari and Roy\cite{roy} that in a unified theory precise bounds
can be
obtained by requiring that the couplings be perturbative up to the
unification scale.  They looked at only the $\la_{dbt}$ and
$\la_{sbt}$ couplings, although their results can easily be
generalized to all of the others.  Specializing to real couplings, the
renormalization group
equations for the $\la_{ijk}$ are given by
\beq \frac{1}{2\pi}{\partial\over\partial t}\la_{ijk}=\gamma_i^a\la_{ajk}
+\gamma_j^b\la_{ibk}+\gamma_k^c\la_{ijc}\eeq where the
$\gamma^a_i=(Z^d_i)^{-1/2}\partial/\partial t(Z^a_d)^{1/2}$ and $Z^a_i$
relates the renormalized superfield $\Phi^i$ to the unrenormalized
$\Phi^a_o$.  For example, the renormalization group equation for
$\la_{dbt}$ is \beq \frac{1}{2\pi} {\partial\over\partial t}\la_{dbt}=
\la_{dbt}(\gamma^d_d+\gamma^b_b+\gamma_t^t)+\la_{dst}\gamma^s_b
+\la_{sbt}\gamma^s_d+\la_{dbu}\gamma^u_t+\la_{dbc}\gamma_t^c.\eeq
The anomalous dimensions can be obtained from the formulae listed by
Martin and Vaughn\cite{vaughn}.  The diagonal $\gamma$'s are given
by\footnote{The factors of 2 in front of the $\la$ terms were given as
$6$ in Ref.  7. We thank Herbi Dreiner for pointing out the correct
expressions.}
\beq 16\pi^2\gamma_d^d=2(\la_{dbt}^2+\la_{dbc}^2+\la^2_{dbu}+\la^2_{dst}+
\la^2_{dsc}+\la^2_{dsu})-{8\over 3}g^2_s-{2\over 15}g^{\prime 2}
\eeq
 \beq16\pi^2\gamma_u^u=2(\la_{dbu}^2+\la_{dsu}^2+\la^2_{sbu})-{8\over
3}g^2_s-{8\over 15}g^{\prime 2}\eeq for the $u$ and $d$ superfields; the
generalization to the other generations is obvious by permutation.
There is an additional term   $2h^2_t$ on the right hand
side of the expression for $16 \pi^2 \gamma_t^t$  due to the top quark Yukawa
coupling. The off-diagonal $\gamma$'s
are given, for example, by \beq
16\pi^2\gamma^d_s=2(\la_{dbt}\la_{sbt}+\la_{dbc}\la_{sbc}+\la_{dbu}\la_{
sbu})\eeq
\beq
16\pi^2\gamma^u_c=2(\la_{dsu}\la_{dsc}+\la_{dbu}\la_{dbc}+\la_{sbu}\la_{
sbc})\eeq

If we define $Y_3\equiv(\la^2_{dst}+\la^2_{dbt}+\la^2_{sbt})/4\pi$,
$Y_2\equiv(\la^2_{dsc}+\la^2_{dbc}+\la^2_{sbc})/4\pi$ and
$Y_1\equiv(\la^2_{dsu}+\la^2_{dbu}+\la^2_{sbu})/4\pi$, then the
renormalization group equations can be combined to yield equations for
$Y_1$, $Y_2$ and $Y_3$. These equations, which in general cannot be written in
terms of the $Y_i$'s alone, take a particularly simple form when  one of the
$Y_i$'s  dominates
  (e.g., when the couplings in $Y_1$ and $Y_2$ do not
significantly contribute to the beta-function for $Y_3$), we have\footnote
{We ignore the hypercharge term, since it is clearly smaller
than the uncalculated two-loop strong interaction corrections.}
\beq  {\partial Y_i\over \partial
t}=6\,Y_i^2-8\,\alpha_s\, Y_i+2\,\delta_{i3}\,h_t^2\,Y_i.\eeq
We now simply require that
the $Y_i$ not become nonperturbative (i.e. $Y_i<1$) by the unification
scale; this leads to a bound at low energies on the $Y_i$.  The
renormalization group equations for $Y_1$, $Y_2$ and (if the top quark
Yukawa coupling is small) $Y_3$, using $\alpha_s$ as it results from the one
loop beta function, can be solved exactly yielding \beq
  Y(\mu)=\alpha_s^{\frac{8}{3}}(\mu)/\left(
\frac{6}{5}\alpha_s^{\frac{5}{3}}(\mu)+C_Y\right)
\eeq where   we replaced $t=\frac{1}{2\pi} \log(\frac{\mu}{M_W})$, and $C_Y$ is
given in terms of $Y  (\mu=M_W)$
\beq C_Y= \alpha_s^{\frac{8}{3}}(M_W)\left(\frac{1}{Y(M_W)}-
\frac{6}{5}\alpha_s^{-1}(M_W)\right).\eeq
For
$\alpha_s(M_W)=0.125$, the requirement of perturbative unification yields
a bound of
$Y_i(M_W)<0.124$ giving an upper bound of $1.25$ on
$\la_{dsu},\la_{dbu},\la_{sbu},\la_{dsc},\la_{dbc}$ and $\la_{sbc}$.
 The bound on $\la_{dst},\la_{dbt}$ and $\la_{sbt}$ does
depend on the top quark Yukawa coupling and must be integrated
numerically.  As shown by Brahmachari and Roy, however, the resulting
bound is very insensitive to $\tan\beta$ and is also insensitive to the
top quark mass--using a top quark mass below about 180 GeV changes the
upper bound by less than ten percent, and using a heavier top quark
causes the top quark Yukawa coupling to become nonperturbative by the
unification scale. It is  worthwhile to mention that $Y_i(\mu)$ decreases
up to values of $\mu$ around  $10^7$ GeV, an effect due to  the running of
$\alpha_s$. Note that the first two couplings are bounded already
from neutron oscillations or nuclear decay, but the bound of
approximately $1.25$ is the strongest bound on the remaining seven.
If one relaxes the assumption that only one $Y_i$  dominates, it is
easy to show that the bounds, in all cases, become strengthened. This is
because all the new contributions to the RHS of equation (17)
are positive.

The bounds we have obtained are on the individual couplings.  On the other
hand, bounds can be obtained on the products of various couplings.  These
bounds are discussed in detail by Barbieri and Masiero\cite{barb}, who
considered the effects of these interactions  in $K-\ov{K}$ mixing, on
$\epsilon^{\prime}/\epsilon$ and on the neutron electric dipole moment.
The first of these gives bounds on $\la_{sbc}\la_{dbc}$ and on
$\la_{sbt}\la_{dbt}$; the latter two give bounds on the imaginary part of the
product of two couplings.  The reader is cautioned that these bounds were
obtained for a top quark mass of $45$ GeV, although changing them to
incorporate a heavier top quark is simple.

We thank  Herbi Dreiner for many enlightening discussions and references,  and
Rabi Mohapatra for helpful remarks concerning neutron oscillations.
  J.L.G. was    supported by NSF grant HRD-9154080
and by DOE contract  DE-AC05-84ER40150, and    M.S.
was supported by NSF grant  PHY-9306141.

\def\prd#1#2#3{{\it Phys. ~Rev. ~}{\bf D#1} (19#2) #3}
\def\prc#1#2#3{{\it Phys. ~Rev. ~}{\bf C#1} (19#2) #3}
\def\plb#1#2#3{{\it Phys. ~Lett. ~}{\bf B#1} (19#2) #3}
\def\npb#1#2#3{{\it Nucl. ~Phys. ~}{\bf B#1} (19#2) #3}
\def\npa#1#2#3{{\it Nucl. ~Phys. ~}{\bf A#1} (19#2) #3}
\def\prl#1#2#3{{\it Phys. ~Rev. ~Lett. ~}{\bf #1} (19#2) #3}

\bibliographystyle{unsrt}

\newpage

\begin{center}
\vspace*{0.2in}
\vbox{\centerline{\epsfig{file=deltaBfig2.eps,height=10.0cm}}}
\vspace*{7 mm}
\parbox{4.5in}{{\bf Figure 1:}  {\it Box diagram giving the necessary neutral
flavor
mixing necessary in $n-\ov{n}$ oscillations. Similar diagrams  appear with
$\tilde{b}$ substituted by $\tilde{s}$.}}
\vspace*{7 mm}
\end{center}
\begin{center}
\vspace*{0.2in}
\vbox{\centerline{\epsfig{file=deltaBfig3.eps,height=8.0cm}}}
\vspace*{7 mm}
\parbox{4.5in}{{\bf Figure 2:}  {\it  Diagram mediating $n-\ov{n}$
oscillations. All particles with no index are right handed. The big blob
represents the insertion of the box in Figure 1, and the small blobs in the
$\tilde{b}$ propagators are insertions of  $M_{\tilde{b}_{LR}}$. }}
\vspace*{7 mm}
\end{center}
\begin{center}
\vspace*{0.2in}
\vbox{\centerline{\epsfig{file=deltaBfig4.eps,height=14.0cm}}}
\vspace*{7 mm}
\parbox{4.5in}{{\bf Figure 3:}  {\it  Bound on $|\la_{dbu}|$ from $n-\ov{n}$
oscillations as a function of $M_{\tilde{t}}$. The solid line corresponds to
$M_{\tilde{u}}=M_{\tilde{c}}=200$ {\rm GeV}, the dashed  line to 400  {\rm GeV}
and the dotted line to 600 {\rm GeV}.}}
\vspace*{7 mm}
\end{center}

 \begin{center}
\vspace*{0.2in}
\vbox{\centerline{\epsfig{file=deltaBfig1.eps,height=10.0cm}}}
\vspace*{7 mm}
\parbox{4.5in}{{\bf Figure 4:}  {\it $\Delta B=-2$, $\Delta S=-2$ diagrams
contributing to double nucleon decay. $(q_i,\, q_j,\,  q_k )$ corresponds to
permutations of $(u,\, d, \, s)$, and similarly for the primed quarks.}}
\vspace*{7 mm}
\end{center}

\end{document}